\documentclass[a4paper]{jpconf}
\usepackage{graphicx}
\begin{document}
\title{Neutrino Mass Models: a road map\footnote{Talk at Neutrino '08, the XXIII International Conference on Neutrino Physics and
Astrophysics, Christchurch, New Zealand.}}

\author{S.F.King}

\address{School of Physics and Astronomy, University of Southampton, Southampton SO17~1BJ, UK}

\ead{king@soton.ac.uk}

\begin{abstract}
In this talk we survey some of the recent
promising developments in the search for the theory behind neutrino mass and mixing,
and indeed all fermion masses and mixing. The talk is organized in terms of a neutrino
mass models road map according to which the answers to experimental questions provide sign posts to guide us through the
maze of theoretical models eventually towards a complete theory of flavour and unification.
\end{abstract}

\section{Introduction}
It has been one of the long standing goals of theories of particle physics beyond the
Standard Model (SM) to predict quark and lepton masses and mixings. With the
discovery of neutrino mass and mixing, this quest has received a massive impetus.
Indeed, perhaps the greatest advance in particle physics
over the past decade has been the discovery of
neutrino mass and mixing involving two large mixing angles commonly known as the
atmospheric angle $\theta_{23}$ and the solar angle $\theta_{12}$, while the
remaining mixing angle $\theta_{13}$, although unmeasured, is constrained to be relatively small
\cite{1}. The largeness of the two large lepton mixing angles contrasts sharply with
the smallness of the quark mixing angles, and
this observation, together with the smallness
of neutrino masses, provides new and tantalizing clues in the search for the origin
of quark and lepton flavour. However, before trying to address such questions,
it is worth recalling why neutrino mass forces us to go beyond the SM.

\section{Why go beyond the Standard Model?}
Neutrino mass is zero in the SM for three independent
reasons:
\begin{enumerate}
\item There are no right-handed neutrinos $\nu_R$.
\item There are only Higgs doublets of $SU(2)_L$.
\item There are only renormalizable terms.
\end{enumerate}
In the SM these conditions all apply and so neutrinos are massless
with $\nu_e$, $\nu_{\mu}$, $\nu_{\tau}$ distinguished by separate
lepton numbers $L_e$, $L_{\mu}$, $L_{\tau}$. Neutrinos and
antineutrinos are distinguished by total conserved lepton number
$L=L_e+L_{\mu}+L_{\tau}$. To generate neutrino mass we must relax
one or more of these conditions. For example, by adding
right-handed neutrinos the Higgs mechanism of the Standard Model
can give neutrinos the same type of mass as the electron mass or
other charged lepton and quark masses. It is clear that the {\it status quo} of staying
within the SM, as it is usually defined, is not an option, but in what direction
should we go?

\section{A road map}
\begin{figure}[h]
\begin{center}
\includegraphics[width=26pc]{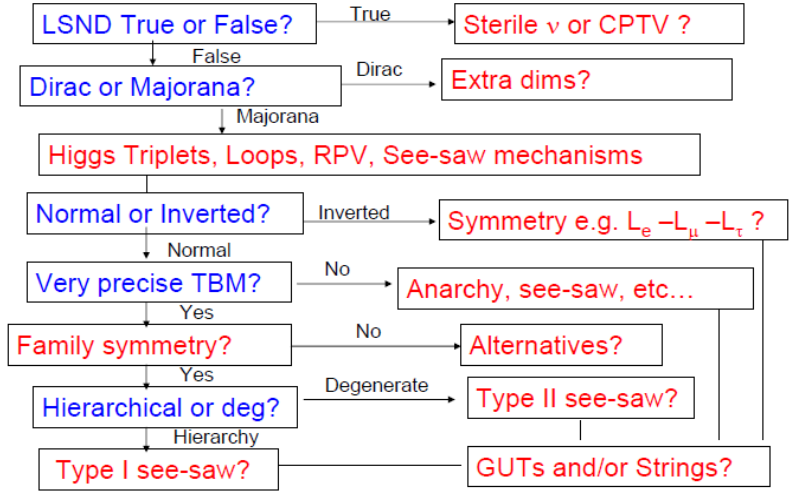}
\caption{\label{roadmap}Neutrino mass models roadmap.}
\end{center}
\end{figure}

This talk will be organized according to the road map
in Fig.\ref{roadmap}. Such a road map is clearly not unique (everyone can
come up with her or his personal road map).
The road map in Fig.\ref{roadmap} contains key experimental
questions (in blue) which serve as signposts along the way, leading
in particular theoretical directions,
starting from the top left hand corner with the question ``LSND True or False?''

\section{LSND True or False?}
As discussed in \cite{2}, the results from MiniBOONE do not support the LSND result,
but are consistent with the three active neutrino oscillation paradigm. If LSND were correct
then this could imply either sterile neutrinos and/or CPT violation, or something more exotic.
For the remainder of this talk we shall assume that LSND is false, and
focus on models without sterile neutrinos. Sterile neutrinos
were discussed in this conference in \cite{3}.

\section{Dirac or Majorana?}
Majorana neutrino masses are of the form $m_{LL}^{\nu}\overline{\nu_L}\nu_L^c$
where $\nu_L$ is a left-handed neutrino field and $\nu_L^c$ is
the CP conjugate of a left-handed neutrino field, in other words
a right-handed antineutrino field. Such Majorana masses are possible
since both the neutrino and the antineutrino
are electrically neutral. Such Majorana neutrino masses
violate total lepton number $L$ conservation, so the neutrino
is equal to its own antiparticle.
If we introduce right-handed neutrino fields then there are two sorts
of additional neutrino mass terms that are possible. There are
additional Majorana masses of the form
$M_{RR}^{\nu}\overline{\nu_R}\nu_R^c$.
In addition there are
Dirac masses of the form
$m_{LR}^{\nu}\overline{\nu_L}\nu_R$.
Such Dirac mass terms conserve total lepton number $L$, but violate
separate lepton numbers $L_e, L_{\mu}, L_{\tau}$.
The question of ``Dirac or Majorana?'' is a key experimental question which could
be decided by the experiments which measure neutrino masses directly \cite{4}.

\section{What if Neutrinos are Dirac?}
Introducing right-handed neutrinos $\nu_R$ into the SM (with zero Majorana mass) we
can generate a Dirac neutrino mass from a coupling to the Higgs:
$\lambda_{\nu} <H>\overline{\nu_L}\nu_R \equiv m_{LR}^{\nu}\overline{\nu_L}\nu_R$,
where $<H>\approx  175$ GeV is the Higgs vacuum expectation value (VEV).
A physical neutrino mass of
$m_{LR}^{\nu}\approx 0.2$ eV implies $\lambda_{\nu}\approx 10^{-12}$.
The question is why are such neutrino Yukawa couplings so small,
even compared to the charged fermion Yukawa couplings?
One possibility for small Dirac masses comes from the idea of extra dimensions
motivated by theoretical attempts to extend the Standard Model
to include gravity .

For the case of ``flat'' extra dimensions,
``compactified'' on circles
of small radius $R$ so that they are not normally observable,
it has been
suggested that right-handed neutrinos (but not the rest of the Standard
Model particles) experience one or more of these extra dimensions \cite{Antoniadis:2005aq}.
%The right handed neutrinos then only spend part of their time in our world,
%leading to very small Dirac neutrino masses
%\cite{Arkani-Hamed:1998vp}.
%In such theories there is a relation between the usual four dimensional Planck mass
%$M_{Planck}\sim 10^{19}\ GeV/c^2$, the string scale $M_{string}$ and the compactification
%radius of the ``flat'' extra dimensions $R$ given by:
%\begin{equation}
%M_{Planck}^2=M_{string}^{2+n}R^n
%\label{flat}
%\end{equation}
%where there are $n$ extra dimensions.
For example, for one extra dimension the right-handed neutrino
wavefunction spreads out over the extra dimension $R$, leading to
a suppressed Higgs interaction with the left-handed neutrino.
%with a suppression factor of $1/\sqrt{M_{string}R}$.
%This corresponds to the coupling between left and right-handed neutrinos
%being more suppressed
%for larger $R$, as the right-handed neutrino spends less
%of its time on the 3 space dimensional brane where the left-handed
%neutrino lives the larger $R$ becomes.
The Dirac neutrino mass
is therefore suppressed relative to the electron mass, and may
be estimated as:
%\begin{equation}
$m_{LR}^{\nu}
%\sim \frac{m_e}{\sqrt{M_{string}R}}
\sim \frac{M_{string}}{M_{Planck}} m_e$
%\end{equation}
%where we have used Eq.\ref{flat} with $n=1$.
where $M_{Planck}\sim 10^{19}\ GeV/c^2$, and $M_{string}$ is the string scale.
Clearly low string scales,
below the Planck scale, can lead to suppressed Dirac neutrino masses.
Similar suppressions can be achieved with anisotropic
compactifications \cite{Antusch:2005kf}.

For the case of ``warped'' extra dimensions things are more complicated/interesting \cite{Huber:2002sp}.
Typically there are two branes, a ``Planck brane'' and a ``TeV brane'', with all the
fermions and the Higgs in the ``bulk'' and having different ``wavefunctions''
which are more or less strongly peaked on the TeV brane. The strength of the Yukawa coupling
to the Higgs is determined by the overlap of a particular fermion wavefunction with the
Higgs wavefunction, leading to exponentially suppressed Dirac masses.
For example the Higgs and
top quark wavefunctions are both strongly peaked on the TeV brane, leading to a large
top quark mass, while the neutrino wavefunctions will be strongly peaked on the
Planck brane leading to exponentialy suppressed Dirac masses.

\section{What if Neutrinos are Majorana?}
We have already remarked that neutrinos, being electrically
neutral, allow the possibility of Majorana neutrino masses.
However such masses are forbidden in the SM since
neutrinos form part of a lepton doublet $L$,
and the Higgs field also forms a doublet $H$, and $SU(2)_L\times U(1)_Y$
gauge invariance forbids a Yukawa interaction like $HLL$. So, if we want
to obtain Majorana masses, we must go beyond the SM.

One possibility is to
introduce Higgs triplets $\Delta$ such that a Yukawa interaction like $\Delta LL$
is allowed. However the limit from the SM $\rho$ parameter implies that the
Higgs triplet should have a VEV $<\Delta > < 8$ GeV. One big advantage is that
the Higgs triplets may be discovered at the LHC and so
this mechanism of neutrino mass generation is directly testable \cite{Perez:2008ha}.

Another possibility, originally suggested by Weinberg,
is that neutrino Majorana masses originate
from operators $HHLL$ involving two Higgs doublets and two lepton doublets,
which, being higher order, must be suppressed by some large mass scale(s)
$M$. When the Higgs doublets get their VEVs Majorana neutrino masses result:
%$\frac{\lambda_{\nu}}{M}<H>^2LL \rightarrow  m_{LL}^{\nu}\overline{\nu_L}\nu_L^c$
$m_{LL}^{\nu}=\lambda_{\nu}<H>^2/M$. This is nice because the
large Higgs VEV $<H>\approx  175$ GeV can lead to small neutrino
masses providing that the mass scale $M$ is high enough. E.g. if
$M$ is equal to the GUT scale $1.75.10^{16}$ GeV then
$m_{LL}^{\nu}=\lambda_{\nu}1.75.10^{-3}$ eV. To obtain larger
neutrino masses we need to reduce $M$ below the GUT scale (since
we cannot make $\lambda_{\nu}$ too large otherwise it becomes
non-perturbative).

Typically in physics whenever we see a large mass scale $M$ associated with a non-renormalizable
operator we tend to associate it with tree level exchange of some heavy particle or particles of mass $M$
in order to make the high energy theory renormalizable once again.
This idea leads directly to the see-saw mechanism where the exchanged particles can either couple
to $HL$, in which case they must be either fermionic singlets (right-handed neutrinos) or
fermionic triplets,
or they can couple to $LL$ and $HH$, in which case they must be scalar triplets.
These three possibilities have been called the type I, III and II see-saw mechanisms, respectively.
%The type I and type II see-saw mechanisms are illustrated in Figs.\ref{I},\ref{II}.
%\begin{figure}[h]
%\begin{minipage}{10pc}
%\includegraphics[width=10pc]{typeI.eps}
%\caption{\label{I}Type I see-saw mechanism.}
%\end{minipage}\hspace{6pc}%
%\begin{minipage}{10pc}
%\includegraphics[width=10pc]{typeII.eps}
%\caption{\label{II}Type II see-saw mechanism.}
%\end{minipage}
%\end{figure}
If the coupling $\lambda_{\nu}$ is very small
(for some reason) then $M$ could even be lowered to the TeV scale
and the see-saw scale could be probed at the LHC \cite{King:2004cx}, however the
see-saw mechanism then no longer solves the problem of the
smallness of neutrino masses.

There are other ways to generate Majorana neutrino masses which lie outside of the above discussion.
One possibility is to introduce additional Higgs singlets and triplets in such a way as to allow neutrino
Majorana masses to be generated at either one \cite{Zee:1980ai} or two \cite{Babu:1988ki} loops.
Another possibility is within the framework of R-parity violating Supersymmetry in which the
sneutrinos $\tilde{\nu}$ get small VEVs inducing a mixing between neutrinos and neutralinos
$\chi$
%via diagrams
%similar to Fig.\ref{I}, but with sneutrinos replacing the Higgs and neutralinos replacing the
%right-handed neutrinos,
leading to Majorana neutrino masses $m_{LL}\approx <\tilde{\nu}>^2/M_{\chi}$,
where for example $<\tilde{\nu}>\approx $ MeV, $M_{\chi}\approx $ TeV leads to $m_{LL}\approx $ eV.
A viable spectrum of neutrino masses and mixings can be achieved at the one loop level \cite{Hirsch:2000ef}.

\section{Normal or Inverted?}
If the mass ordering is inverted (as defined for example in \cite{5}) then
this may indicate a new symmetry such as $L_e - L_{\mu} - L_{\tau}$ \cite{Petcov:1982ya}
or a $U(1)$ family symmetry \cite{King:2000ce}. However let us assume that the hierarchy is
normal and proceed down the road map to the next experimental question.

\section{Very precise tri-bimaximal mixing?}
It is a striking fact that current data on lepton mixing is consistent with
the so-called tri-bimaximal (TB) mixing pattern \cite{Harrison:2002er},
\begin{equation}
\label{TBM}
U_{TB}= \left(\begin{array}{ccc} \sqrt{\frac{2}{3}}& \frac{1}{\sqrt{3}}&0\\
-\frac{1}{\sqrt{6}}&\frac{1}{\sqrt{3}}&\frac{1}{\sqrt{2}}\\
\frac{1}{\sqrt{6}}&-\frac{1}{\sqrt{3}}&\frac{1}{\sqrt{2}} \end{array} \right)
P_{Maj},
\end{equation}
where $P_{Maj}$ is the diagonal phase matrix involving the two observable
Majorana phases. However there is no convincing reason to expect exact TB mixing, and
in general we expect deviations. These deviations can be parametrized by three parameters
$r,s,a$ defined as \cite{King:2007pr}:
%\begin{equation}
$\sin \theta_{13} = \frac{r}{\sqrt{2}}$, $\sin \theta_{12} = \frac{1}{\sqrt{3}}(1+s)$,
$\sin \theta_{23} = \frac{1}{\sqrt{2}}(1+a)$.
%\label{rsa}
%\end{equation}
Global fits of the conventional mixing angles
\cite{Schwetz:2008er} can be translated into the $2\sigma$ ranges
\footnote{Note that $r$ must be positive definite,
while $s,a$ can take either sign. Indeed there is a preference for $s$ to be negative.}
%\begin{equation}
$0<r<0.28$, $-0.10<s<0.02$, $-0.12<a<0.12$.
%\end{equation}
Very precise TB mixing would correspond to $r,|s|,|a|\ll 1$, and
TB mixing would demand an explanation,
while if $r,s,a$ are close to their current $2\sigma$ bounds, then TB mixing
would only be realized approximately and could be just a coincidence.
But the question is how small would $r,|s|,|a|$ have to be in order for TB mixing not to be
a coincidence? This question has been addressed in \cite{Albright:2008qb} where it is argued that
the crucial parameter is $U_{e3}$ or $r$, and that if this parameter is much smaller than about 0.1
then it would be hard to argue that TB mixing is accidental. On the other hand if $r>0.1$ then perhaps
we should not regard the angle $\theta_{13}$ as being particularly small, and one possibility is that
all the lepton mixing angles are chosen at random, for example as in ``Anarchy'' \cite{Hall:1999sn}.

\section{Family Symmetry?}
Assuming that TB mixing is very precise and is not an accident, it could be interpreted as a signal
of an underlying family symmetry. Indeed I am unaware of any viable alternative at present.
To understand the emergence of a family symmetry, let us expand the
neutrino mass matrix in the diagonal charged lepton basis, assuming exact TB mixing,
as $m_{LL}^{\nu}=U_{TB}{\rm diag}(m_1, m_2, m_3)U_{TB}^T$ leading to (absorbing the Majorana phases in $m_i$):
\begin{equation}
\label{mLL}
m_{LL}^{\nu}=\frac{m_3}{2}\Phi_3 \Phi_3^T + \frac{m_2}{3}\Phi_2 \Phi_2^T + \frac{m_1}{6}\Phi_1 \Phi_1^T
\end{equation}
where $\Phi_3^T=(0,1,-1)$, $\Phi_2^T=(1,1,1)$, $\Phi_1^T=(2,-1,1)$ and $m_i$ are the physical neutrino masses.
This shows that the neutrino mass matrix corresponding to TB mixing may be constructed from
the very simple orthogonal column vectors $\Phi_i$, whose simplicity motivates an underlying
non-Abelian family symmetry involving all three families. The idea is that $\Phi_i$ are promoted to
new Higgs fields called ``flavons'' whose VEVs break the family symmetry, with the particular
vacuum alignments as above. Such vacuum alignments can more readily be achieved if the non-Abelian family symmetry
is a discrete symmetry containing a permutation symmetry capable of leading to $<\Phi_2^T>\propto (1,1,1)$
\cite{Altarelli:2005yp}.
A minimal choice of such family symmetry seems to be $A_4$ \cite{Ma:2001dn}
which only involves the flavon $<\Phi_2^T>\propto (1,1,1)$
together with a further flavon $<\Phi_0^T>\propto (0,0,1)$. Such minimal $A_4$ models lead to neutrino mass sum rules
between the three masses $m_i$, resulting in/from a simplified mass matrix in Eq.\ref{mLL}.
$A_4$ may result from 6D orbifold models \cite{Altarelli:2006kg}.

It is possible to derive the TB form of the neutrino mass matrix in Eq.\ref{mLL} from the see-saw
mechanism in a very elegant way using the idea of constrained sequential dominance (CSD)\cite{King:2005bj}
as follows. In the diagonal right-handed neutrino mass basis we may write $M_{RR}^{\nu}={\rm diag}(M_A, M_B, M_C)$
and the Dirac mass matrix as $m_{LR}^{\nu}=(A,B,C)$ where $A,B,C$ are three column vectors. Then the
type I see-saw formula $m_{LL}^{\nu}=m_{LR}^{\nu}(M_{RR}^{\nu})^{-1}(m_{LR}^{\nu})^T$ gives
\begin{equation}
\label{mLLCSD}
m_{LL}^{\nu}=\frac{AA^T}{M_A}+ \frac{BB^T}{M_B} + \frac{CC^T}{M_C}.
\end{equation}
By comparing Eq.\ref{mLLCSD} to the TB form in Eq.\ref{mLL} it is clear that
TB mixing will be achieved if $A\propto \Phi_3$, $B\propto \Phi_2$, $C\propto \Phi_1$,
with each of $m_{3,2,1}$ originating from a particular
right-handed neutrino of mass $M_{A,B,C}$, respectively \cite{King:2005bj}.
If $m_1\ll m_2 < m_3$ then the precise form of $C$ becomes irrelevant, and the
CSD mechanism has been applied in this case to models based on the family symmetries $SO(3)$
\cite{King:2005bj,King:2006me} and
$SU(3)$ \cite{deMedeirosVarzielas:2005ax}, and their discrete subgroups \cite{deMedeirosVarzielas:2005qg}.

\section{Hierarchical or Degenerate?}
This key experimental question may be decided by the same experiments as will also determine
the nature of neutrino mass (Dirac or Majorana) \cite{4}.
Although not a theorem, it seems that a hierarchical spectrum could indicate a type I see-saw mechanism,
while a (quasi) degenerate spectrum could imply a type II see-saw mechanism. It is possible that a
type II see-saw mechanism could naturally explain the degenerate mass scale with the
degeneracy enforced by an $SO(3)$ family symmetry, while the type I see-saw part could be responsible
for the small neutrino mass splittings and the (TB) mixing \cite{Antusch:2004xd}.

\section{GUTs and/or Strings?}
Finally we have reached the end of the road map, with the possibility
of an all-encompassing unified theory of flavour based on GUTs and/or strings. Such theories could also include
a family symmetry in order to account for the TB mixing. There are many possibilities
for the choice of family symmetry and GUT symmetry.
Examples include the Pati-Salam gauge group
$SU(4)_{PS}\times SU(2)_L\times SU(2)_R$ in combination with $SU(3)$ \cite{deMedeirosVarzielas:2005ax},
$SO(3)$ \cite{King:2005bj,King:2006me}, $A_4$ \cite{King:2006np} or
$\Delta_{27}$ \cite{deMedeirosVarzielas:2006fc}. Other examples are based on $SU(5)$ GUTs
in combination with $A_4$ \cite{Altarelli:2008bg} or $T'$ \cite{Chen:2007afa}.

In typical Family Symmetry $\otimes$ GUT models the origin of the quark mixing angles
derives predominantly from the down quark sector, which in turn is
closely related to the charged lepton sector. In order to
reconcile the down quark and charged lepton masses, simple
ansatze, such as the Georgi-Jarlskog hypothesis \cite{Georgi:1979df}, lead to very
simple approximate expectations for the charged lepton mixing
angles such as $\theta^e_{12}\approx \lambda/3$,
$\theta^e_{23}\approx \lambda^2$, $\theta^e_{13}\approx
\lambda^3$, where $\lambda \approx 0.22$ is the Wolfenstein
parameter from the quark mixing matrix. If the family symmetry
enforces accurate TB mixing in the neutrino sector, then $\theta^e_{12}\approx \lambda/3$
charged lepton corrections will cause deviations from TB mixing in the physical
lepton mixing angles, and lead to a sum rule relation
\cite{King:2005bj,Masina:2005hf,Antusch:2005kw},
which can be conveniently expressed as \cite{King:2007pr}
$s\approx r \cos \delta$ where $r\approx \lambda /3$ and $\delta$ is the observable CP violating
oscillation phase, with RG corrections of less than one degree \cite{Boudjemaa:2008jf}.
Such sum rules can be tested in future high precision neutrino oscillation
experiments \cite{Antusch:2007rk}.

%\begin{figure}[h]
%\begin{minipage}{14pc}
%\includegraphics[width=14pc]{family_groups.eps}
%\caption{\label{family}Some possible family symmetry groups.}
%\end{minipage}\hspace{2pc}%
%\begin{minipage}{14pc}
%\includegraphics[width=14pc]{gut_groups.eps}
%\caption{\label{GUT}Some possible GUT groups.}
%\end{minipage}
%\end{figure}

\section{Conclusion}
Neutrino mass and mixing clearly requires new physics beyond the SM, but in which direction
should we go? There are many roads for model building, but we have seen that answers to key experimental
questions will provide the sign posts {\it en route} to a unified theory of flavour.

\end{document}